\documentclass[letter]{aa}     

\usepackage{graphicx}
\usepackage{txfonts}
\usepackage{lipsum}
\usepackage{subcaption}         
\usepackage{lscape}             
\usepackage{placeins}           
\usepackage{xcolor}               
\usepackage{siunitx}

\begin{document}


%

   \title{Shape and spin axis determination of the Tianwen-2 target asteroid (469219) Kamo'oalewa from lightcurve inversion}
   \titlerunning{Shape and spin axis of asteroid (469219) Kamo'oalewa}

   \author{
     R. Bonamico\inst{1} \and 
     J. Hanu\v{s}\inst{2} \and
     M. Delbo\inst{3,4}
   }

   \institute{
     BSA Osservatorio (K76), Strada Collarelle 53, 12038 Savigliano, Cuneo, Italy
     \and
     Charles University, Faculty of Mathematics and Physics, Institute of Astronomy, V Hole\v{s}ovi\v{c}k\'ach 2, 180\,00 Prague, Czech Republic
     \and
     Laboratoire Lagrange, Centre National de la Recherche Scientifique, Observatoire de la Côte d'Azur, Université Côte d'Azur, 06304Nice, France.
     \and
     University of Leicester, School of Physics and Astronomy, University Road, LE1 7RH, Leicester, UK.\\
     \email{delbo@oca.eu}
     }

   \date{Received March 30, 2026}

 
  \abstract
   {Near-Earth asteroid (469219) Kamo'oalewa is an Earth quasi-satellite, temporarily trapped in a 1:1 orbital resonance with our planet. Despite its dynamical relevance and the hypothesis that it may be a lunar ejecta fragment, its physical properties are still poorly constrained. In particular, no reliable models of its shape and spin state have been published so far. The scientific interest in this object is further enhanced by its selection as the primary target of the Chinese Tianwen-2 mission, which aims to rendezvous with this asteroid and return samples of it to Earth. }
   {The aim of this work is to determine the shape and spin axis orientation of Kamo'oalewa by means of photometric telescope observations and lightcurve inversion.}
   {We analyzed lightcurves obtained during several apparitions using the well-established algorithm, based on convex shape modeling. }
   {We derived a convex shape model and estimated the spin pole orientation. In the preferred solution, the pole is located at ecliptic coordinates {$\lambda$}, {$\beta$}  =  (126, $-$16)$^\circ$, with a sidereal rotation period of P = 0.465 h.}
   {Our results provide the first direct constraints on the rotational state and morphology of Kamo'oalewa, information of key importance in preparation for the upcoming Tianwen-2 sample-return mission. }


   \maketitle
\nolinenumbers

\section{Introduction}

Near-Earth Objects (NEOs) are a dynamically diverse population of small bodies with perihelion distance $Q \leq 1.3$ au \citep[see][for a review]{pernaNearEarthObjectsTheir2013}. They are prime targets for space missions \citep[e.g.,][]{binzelObservedSpectralProperties2004,elvis013aste.book...81E}, as demonstrated by rendezvous \citep{veverkaLandingNEARShoemakerSpacecraft2001}, kinetic impactor \citep{chengMomentumTransferDART2023}, and sample return missions \citep[e.g.,][]{fujiwaraRubblePileAsteroidItokawa2006,laurettaOSIRISRExSampleReturn2017,watanabeHayabusa2ArrivesCarbonaceous2019}. NEOs also pose an impact hazard to Earth \citep{Vavilov2025,Nugent2025}. Near-Earth asteroids (NEAs) are a subset of NEOs. Earth quasi-satellites are NEAs in a 1:1 mean-motion resonance with Earth. Although not gravitationally bound, they remain in the vicinity of Earth for extended periods, making them attractive targets for observations and potential missions \citep{Connors2011,Wiegert2000}.

Asteroid (469219) Kamo'oalewa, discovered in 2016 by the Pan-STARRS1 survey \citep{chambersPanSTARRS1Surveys2019}, is the most stable known Earth quasi-satellite. The absolute magnitude value\footnote{mp3c.le.ac.uk;  version 3.0.0-beta.22 of 2026-03-06} of  24.3, results in a diameter of $\sim$30–100 m for albedos in a range between 0.35 and 0.034. Its orbit and size make it one of the smallest NEAs currently reachable by spacecraft.

Light-curve observations indicate a rotation period of about 28 minutes \citep{sharkeyLunarlikeSilicateMaterial2021}. Ground based spectroscopy reveals a reflectance that can be consistent with silicate material, but with unusual reddening compared to typical inner-solar-system asteroids \citep{sharkeyLunarlikeSilicateMaterial2021}. It has been suggested that its reflectance is mostly compatible with lunar-like silicate materials, raising the intriguing possibility that it could be lunar ejecta \citep{sharkeyLunarlikeSilicateMaterial2021}. Other Earth coorbital asteroids (minimoons) have been found to have spectral colours consistent with lunar-like materials \citep{bolinDiscoveryCharacterizationMinimoon2025}, while others attribute the spectrum to space-weathered material from an S-complex asteroid captured from the NEA population into an Earth-like orbit \citep{sharkeyLunarlikeSilicateMaterial2021}.

Long-term astrometric monitoring has enabled the detection of the Yarkovsky effect \citep{liuYarkovskyEffectDetection2022}, a force due to asymmetric thermal emission that causes slow changes in orbital semimajor axis \citep{bottkeYARKOVSKYYORPEFFECTS2006}, which provides constraints on physical properties like thermal inertia \citep{fenucciLowThermalConductivity2021,novakovicASTERIAAsteroidThermal2024} and bulk density \citep{fenucciLowThermalConductivity2021}. These constraints can be further refined with knowledge of the asteroid's size, shape, and spin state, which influence its thermal environment and Yarkovsky acceleration \citep[e.g.,][and references therein]{delboThermalInertiaNearEarth2007}. 

The Chinese space mission Tianwen-2, launched on 28 May 2025 
is designed to rendezvous with Kamo'oalewa, collect surface samples, and return them to Earth. The spacecraft is expected to reach the asteroid by July 2026 and return the collected samples by 2027\footnote{Space.com. 2025, China's Tianwen-2 launches to visit near-Earth asteroid Kamo'oalewa, https://www.space.com/, published May 28, 2025. Accessed 2025-09-23
SpaceNews. 2025, China launches Tianwen-2 asteroid sample return mission, https://spacenews.com/, published May 28, 2025. Accessed 2025-09-23}. 
This represents the first attempt to sample a quasi-satellite and constitutes a milestone in planetary science, complementing previous sample return missions: JAXA's Hayabusa \citep{fujiwaraRubblePileAsteroidItokawa2006}, Hayabusa2 \citep{watanabeHayabusa2ArrivesCarbonaceous2019}, and NASA's OSIRIS-REx \citep{laurettaAsteroidBennu2024}. 

The success of Tianwen-2 depends on accurate knowledge of Kamo'oalewa’s physical properties, including shape, spin state, and surface characteristics. Ground-based observations can provide key constraints on these parameters. The case of (101955) Bennu illustrates this approach: radar observations constrained its shape and spin \citep{nolanShapeModelSurface2013}, thermal-infrared data its thermal properties \citep{emeryThermalInfraredObservations2014a}, spectroscopy its composition \citep{clarkSpectroscopyBtypeAsteroids2010}, and photometry its rotation and surface properties \citep{hergenrotherLightcurveColorPhase2013}. Such pre-encounter characterization is essential for NEA missions. 

Light-curve inversion techniques enable the reconstruction of asteroid shape and spin from ground-based photometry \citep{kaasalainenOptimizationMethodsAsteroid2001a,kaasalainenOptimizationMethodsAsteroid2001,durechDAMITDatabaseAsteroid2010}. The method has been extensively validated through comparisons with radar, adaptive optics, stellar occultations, and spacecraft imaging \citep[e.g.,][]{hanusPhysicalCharacterizationAsteroid2025}, and is widely applied, with thousands of models available in the DAMIT database.

Here we present a convex shape model of Kamo'oalewa derived from dense photometric data obtained over multiple apparitions, together with a pole search and rotation-period analysis. Section~2 describes the data, Section~3 the inversion method, Section~4 the results, and Section~5 discusses the implications.

\section{Source of data}


The lightcurves used here for the inversion were based on the photometric data downloaded from the \texttt{AstDys-2} database, which includes photometric observations of Kamo'oalewa from three apparitions in 2016, 2017, and 2018. These data were acquired with the 2.2-meter (88-inch) University of Hawai'i telescope (UH88) on Mauna Kea (IAU code T12).
Standard data reduction procedure was adopted as well as aperture photometry relative to field stars using a G-band filter. Information about the exposure times used are not available from the AstDys.  However, since our work is based on relative photometry, the lack of this information is not affecting our results. The parameters relevant for convex shape modeling are summarized in Table~\ref{tab:photometry}.

Due to the rotation period of the asteroid \citep[$P\sim$ 28.3 min;][]{sharkeyLunarlikeSilicateMaterial2021}, each of the 15 observing sessions carried out at the T12 observatory, lasting between 25 and 60 minutes, covered almost one full rotation or more. Consequently, these 15 sessions yielded 15 complete light curves. The T12 UH88 photometric dataset was therefore treated as dense-in-time photometry and included as such in the light-curve inversion process.

\begin{table}
\caption{Photometric observations of (469219) Kamo'oalewa available in AstDys-2. 
The symbol $N_\mathrm{o}$ indicates the number of individual measurements obtained during each apparition.} 
\label{tab:photometry}      
\centering                         
\begin{tabular}{lccc}        
\hline\hline                 
Apparition & Date range & $N_\mathrm{o}$& Phase angle [$^\circ$]  \\    
\hline                        
2016 & 2016--05 -- 2016--06& 25& 56.5--72.6 \\  
2017 & 2017--03 -- 2017--05& 198& 47.6--64.7 \\  
2018 & 2018--04 -- 2018--04& 20& 43.9--43.9 \\  
\hline                                   
\end{tabular}
\end{table}

\section{Methods}

\begin{figure}
   \centering
   \includegraphics[width=0.48\textwidth]{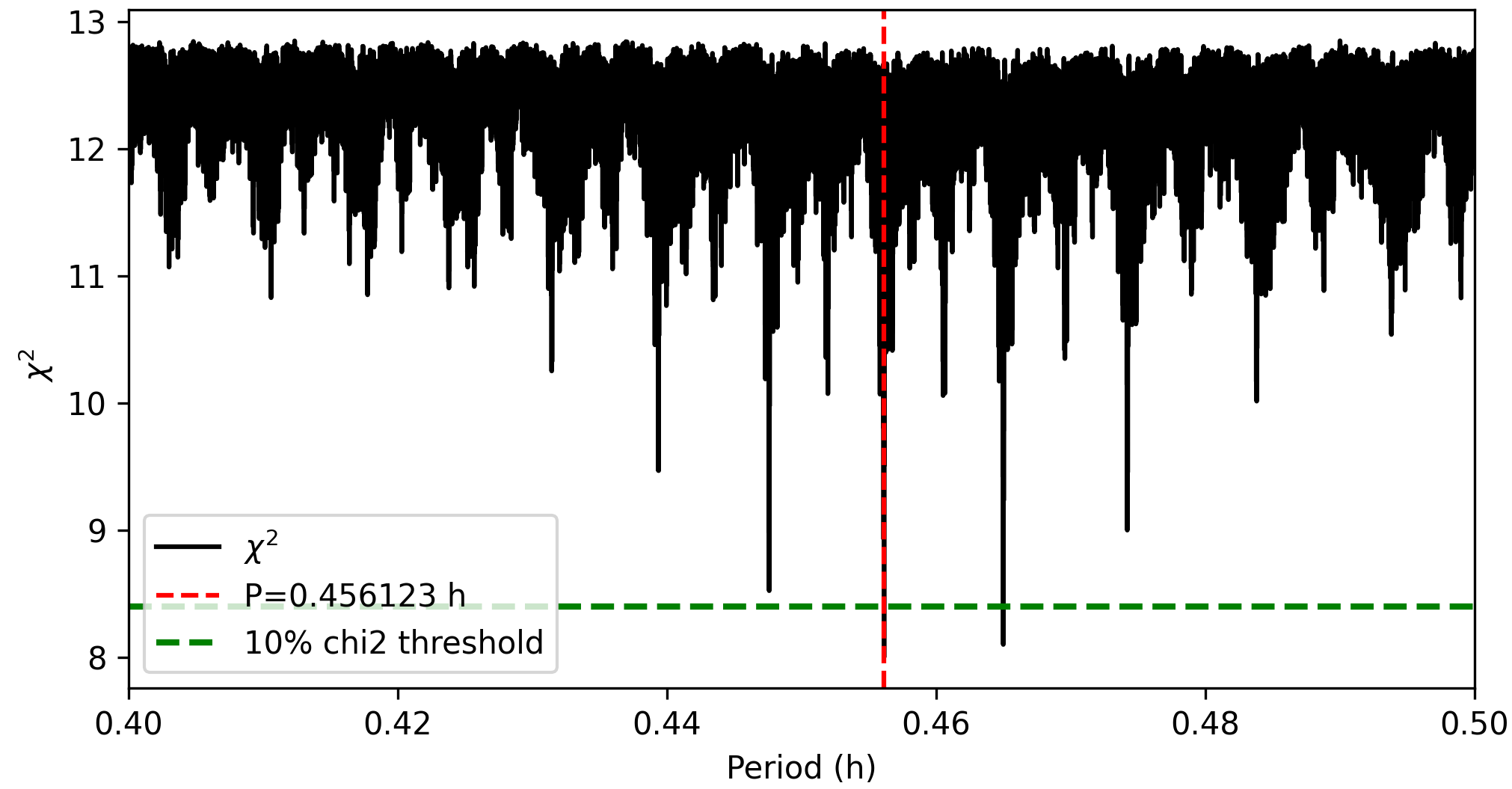}
   \caption{Periodogram -- search for the sidereal rotation period. The horizontal axis shows the tested periods (0.40--0.50 h), and the vertical axis shows the corresponding $\chi^2$ values. Two periods within the $\chi^2$ threshold (defined by the green dashed line) were selected for further analysis.}
   \label{fig:chisq}
\end{figure}


\begin{figure*}
   \centering
   \includegraphics[width=\textwidth]{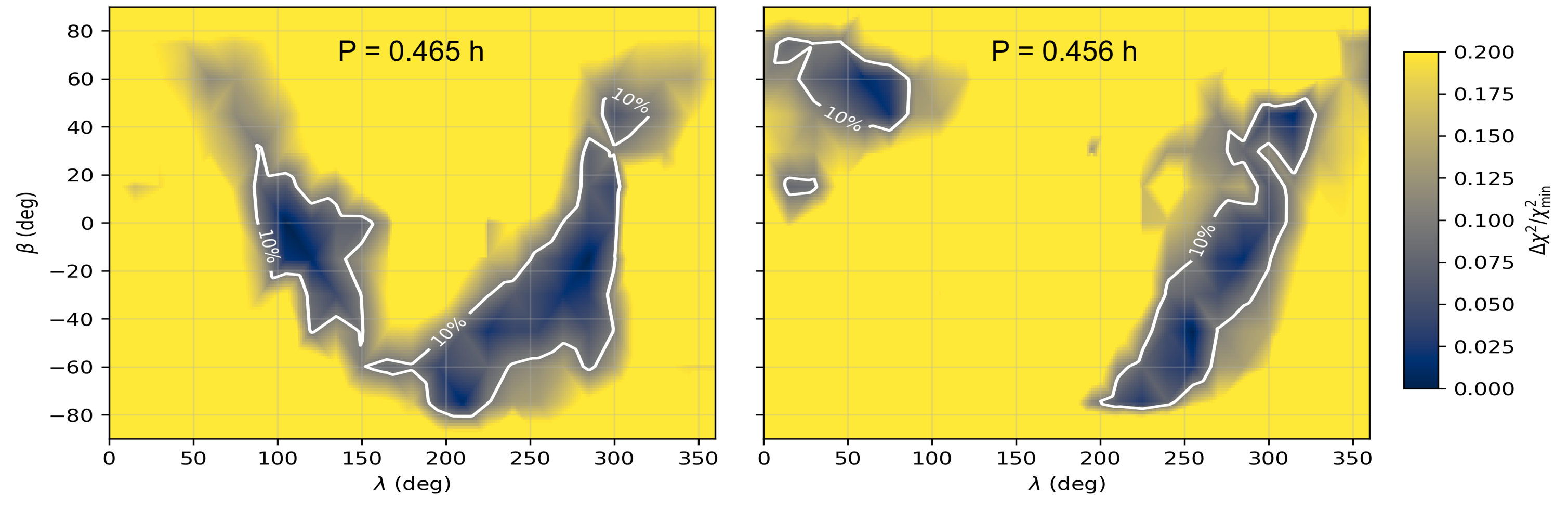}
   \caption{Pole map obtained using the ``medium'' search option of \texttt{MPO LCInvert}. Dark  regions indicate the lowest $\chi^2$ values, while yellow regions indicate the highest. White lines contour the regions of admissible solutions (see Sect.~\ref{s:spin_axis_search}).}
   \label{fig:pole_search}
\end{figure*}

%
%

The light-curve inversion was performed using \texttt{MPO LCInvert} v.11.8.3.1 \citep{Warner2009_LCDB,Warner2012_LCInvert,Warner2006}, implementing the convex inversion method of \citet{kaasalainenOptimizationMethodsAsteroid2001a,kaasalainenOptimizationMethodsAsteroid2001}, later refined by \citet{durechDAMITDatabaseAsteroid2010}. The model assumes a convex polyhedral shape in principal-axis rotation. Surface scattering is described by a linear combination of Lommel–Seeliger and Lambert laws with an empirical phase function \citep{kaasalainenOptimizationMethodsAsteroid2001a}. The shape is parameterized via surface normals, with facet areas optimized through regularized least-squares minimization. The inversion simultaneously solves for the sidereal period, spin-axis orientation, and shape by minimizing $\chi^2$ between observed and synthetic light curves. All photometric data were equally weighted (homogeneous origin). A maximum of 50 iterations per run was sufficient to ensure convergence.

\subsection{Rotation period determination}
\label{s:rot_period_determination}

The first step of the inversion procedure consists of determining the sidereal rotation period. A preliminary Lomb–Scargle analysis indicated a rotation period close to 0.45~h, consistent with the estimate of \citet{sharkeyLunarlikeSilicateMaterial2021}. 

We therefore performed a dense period scan in the interval 0.40–0.50~h using a step of $1\times10^{-5}$~h. For each trial period, the convex inversion was executed and the corresponding $\chi^2$ value recorded. The resulting periodogram is shown in Fig.~\ref{fig:chisq}. 

Instead of a single global minimum, two distinct local minima of comparable depth were identified near 0.465~h and 0.456~h. Both candidate periods fall within the standard acceptance threshold of 10\% above the global $\chi^2$ minimum \citep{Hanus2011}. All other period candidates were rejected. The formal period uncertainty was estimated as one twentieth of the width of the respective $\chi^2$ minima \citep{Hanus2011}, yielding an uncertainty of approximately $1\times10^{-6}$~h for both candidate solutions.

\subsection{Bootstrap validation}
To assess the robustness of the two period candidate solutions, we applied a bootstrap resampling procedure \citep{hanusThermophysicalModelingAsteroids2015}. We generated 700 synthetic datasets by resampling the photometric data with replacement and repeated the period scan over the restricted interval 0.45–0.47~h.

The resulting distribution (Fig.~\ref{fig:bootstrap}) clusters around the two candidate minima, confirming that both are supported by the data. The $P\sim$~0.456~h solution appears in $\sim$69\% of realizations, indicating a statistical preference. However, given the limited number of apparitions and similar $\chi^2$ depths, we retain both periods for further analysis.

\subsection{Spin-axis search}
\label{s:spin_axis_search}

For each candidate rotation period, we performed an exploration of the spin-axis orientation by scanning the celestial sphere in ecliptic coordinates, using a grid with 15$^\circ$ spacing in both ecliptic longitude ($\lambda$) and latitude ($\beta$), corresponding to 312 trial pole directions. For each pole direction, the light-curve inversion was executed with the rotation period fixed to the candidate value, and the corresponding $\chi^2$ recorded.

The resulting $\chi^2$ distribution over the sky reveals regions (dark areas of Fig.~\ref{fig:pole_search}) of admissible solutions rather than isolated points. Following the criterion of \citet{Hanus2011}, we defined all solutions satisfying $\Delta \chi^2 / \chi^2_{\min} < 0.1$, where $\Delta \chi^2 = \chi^2 - \chi^2_{\min}$ as admissible solutions. Not unexpectedly, these solutions are enclosed in regions (white curves of Fig.~\ref{fig:pole_search}) corresponding to local $\chi^2$-minima and represent spin-axis orientations compatible with the data.
This procedure was applied independently for both candidate periods. For the $P=0.465$ h case, two dominant regions were found, forming a pair of mirror pole solutions separated by approximately 180$^\circ$ in longitude. This is the well-known ambiguity of light-curve inversion with limited viewing geometries.

\subsection{Final shape solution}
\label{sec:final_shape_solution}

For each candidate period ($P\sim0.465$~h and $P\sim0.456$~h) starting with the lower $\chi^2$-solution within each separate region of  $\Delta \chi^2 / \chi^2_{\min} < 0.1$ identified by means of the method of Sect.~\ref{s:spin_axis_search}, we performed a refined inversion in which both the rotation period and the spin-axis orientation were allowed to vary simultaneously. This step enables convergence toward the nearest local $\chi^2$ minimum within each region, allowing pole,  period, and shape model to be optimized, and typically yields solutions with slightly lower $\chi^2$ values than those obtained in the fixed-period pole scan. Convergence to one local $ \chi^2$ minimum per region was observed (Table~\ref{tab:spin_solutions}).

\begin{table}
\caption{Solutions at local minima within acceptable regions.} 
\centering
\begin{tabular}{c c c c c c}
\hline\hline
Candidate	& Solution &{$\chi^2$}	&{$\lambda$}	& {$\beta$} 	&{Period}\\
Period (h)	&          	&          		& (deg)     		& (deg)     		& (h) \\
\hline
0.465 & 1 & 7.648 & 126 & $-$16 & 0.464983 \\
      & 2 & 7.822 & 278 & $-$28 & 0.464983 \\
      & 3 & 7.844 & 300 &    45 & 0.464983 \\
\hline
0.456 & 1 & 8.241 & 260 & $-$39 & 0.456123 \\
      & 2 & 8.639 & 73  & 63  & 0.456099 \\
      & 3 & 8.685 & 30 & 15 & 0.456099 \\
\hline
\end{tabular}
%
\tablefoot{Candidate periods were derived using the method of Sect.~\ref{s:rot_period_determination}. Solutions correspond to the local $\chi^2$ minima within each white-contoured region and were obtained using the method of Sect.~\ref{sec:final_shape_solution}.}

\label{tab:spin_solutions}
\end{table}

The resulting set of acceptable solutions, i.e., those within the white curves of Fig.~\ref{fig:pole_search},  defines the uncertainty in the determination of the spin vector and rotation period. This approach gives a realistic estimate of the admissible region in spin-vector space given the data quality and observational geometry.

\section{Results}

The convex inversion analysis reveals two local minima in the explored 
period interval, at $P \sim 0.465$ h and $P \sim 0.456$ h 
(Fig.~\ref{fig:chisq}, Table~\ref{tab:spin_solutions}). 
For $P \sim 0.465$ h, two pole solutions are found at 
$(\lambda, \beta) = (126^\circ, -16^\circ)$ and 
$(278^\circ, -28^\circ)$, with $\chi^2 = 7.648$ and $7.822$, respectively. 
A third, less-preferable solution was also found at $(\lambda, \beta) = (300^\circ, 45^\circ)$ with $\chi^2 = 7.844$.
For $P \sim 0.456$ h, three additional solutions are obtained at 
$(260^\circ, -39^\circ)$, $(73^\circ, 63^\circ)$, and 
$(30^\circ, 15^\circ)$, with $\chi^2$ values between 8.241 and 8.685.
The formally preferred solution corresponds to $P = 0.464982$ h and $(\lambda, \beta) = (126^\circ, -16^\circ)$, providing the lowest $\chi^2$ (7.648). The 
root mean square of the residuals, RMS, between the model and the data is 0.21 magnitudes,  
consistent with the expected photometric scatter (see Appendix~\ref{sec:suppAnal}).

The proximity in $\chi^2$ between the two $P \sim 0.465$ h solutions and their difference in $\lambda$ suggests the typical ambiguity in pole determination arising from limited aspect-angle coverage. Nevertheless, the lowest-$\chi^2$ solution is statistically favored. Considering the extremely short rotation period ($P \sim 0.465$ h), the shape does not exhibit an elongation sufficient to justify rotational stability solely through self-gravity. The object does not show evidence of extreme centrifugal deformation; therefore, its rotational stability is plausibly maintained by non-negligible internal cohesive strength, consistent with expectations for super-fast rotators and with the possibility that it may represent a fragment of lunar bedrock.

Overall, the derived shape model reproduces the observed photometric amplitudes well. The residual scatter of the fit is comparable to the typical photometric uncertainties, indicating that the reconstructed geometry adequately explains the observed brightness variations without requiring unrealistically extreme morphological features.
Figure 2 presents the convex shape model associated with the preferred spin-state solution.

\begin{figure}
   \centering
   \includegraphics[width=0.42\textwidth]{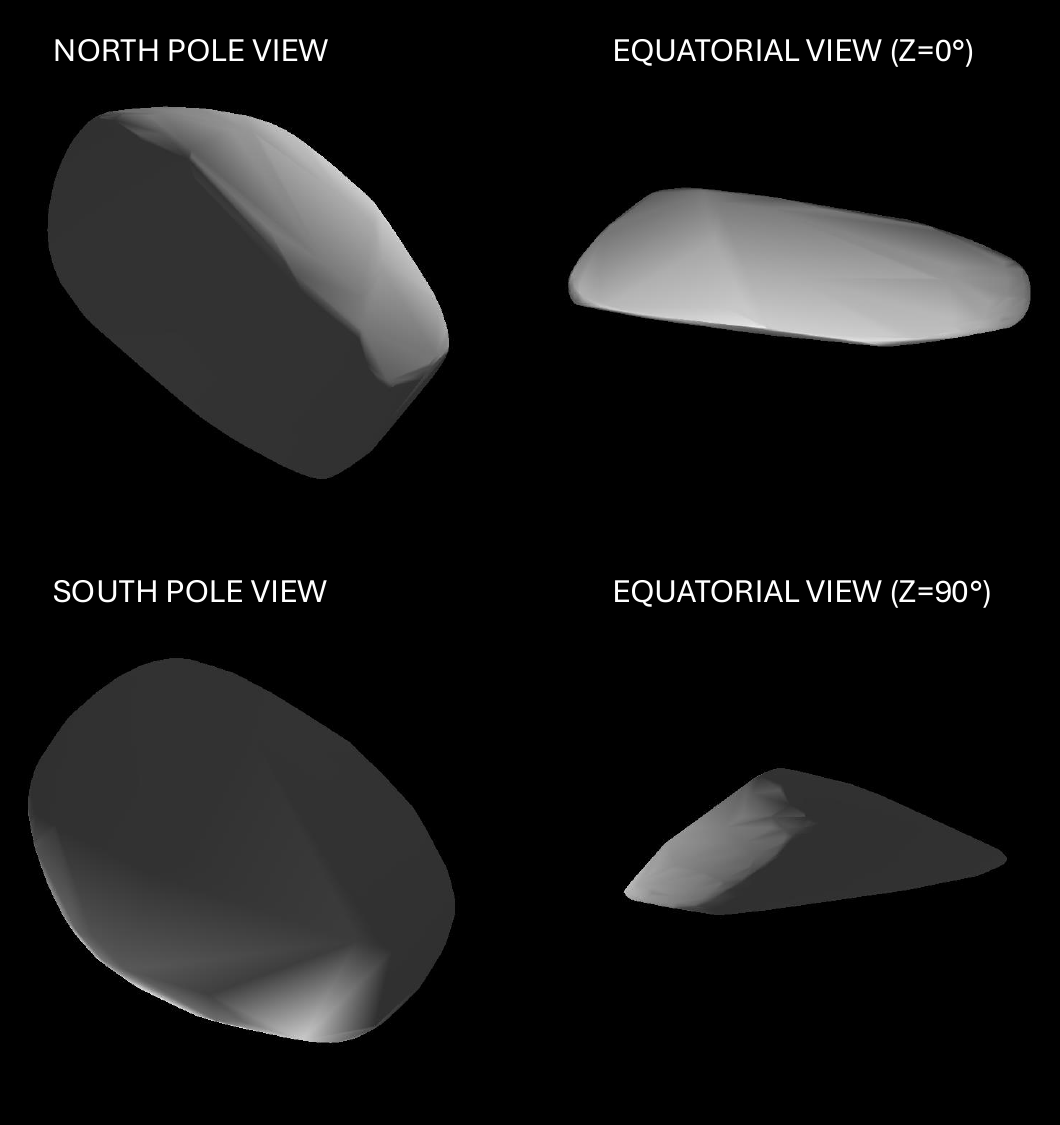}
   \caption{Convex shape model (Solution 1 for the 0.465 h period). }
   \label{fig:Shape Model}
\end{figure}

\section{Discussion}
All alternative solutions (Table~\ref{tab:spin_solutions}) preserve the overall convex and elongated morphology of the preferred model. However, they differ in the distribution of volume along the principal axis. In particular, solution 2 with $P\sim0.465$~h mainly shows a moderate redistribution of volume without significantly altering the global elongation. Solution 3 with $P\sim0.465$~h appears elongated and flattened, with a protrusion at one end. Solution 1 with $P\sim0.456$~h appears more compact and less tapered, with a thicker central region and a more massive extremity; Solution 2 with $P\sim0.456$~h is very similar to the preferred solution. Solution 3 with $P\sim0.456$~h turns out to be flattened and quite angular. None of the models displays bilobate features. For the figures corresponding to all the shapes, see Appendix ~\ref{sec:shape}.

The markedly elongated shape derived for Kamo'oalewa is difficult to reconcile with a purely hydrostatic equilibrium configuration, particularly given its extremely short rotation period. The absence of pronounced equatorial bulging or signatures of centrifugal mass redistribution suggests that the morphology is not the result of rotational deformation of a strengthless body. Instead, it reflects a structure dominated by internal cohesion.

A qualitative comparison with the unusual morphology reported by \citet{Bolin2024YR4} for 2024 YR4 indicates that strongly anisotropic shapes may arise in small bodies as a consequence of collisional fragmentation followed by rotational evolution. In this framework, the observed elongation of Kamo'oalewa may largely preserve the geometry of the original fragment, maintained by non-negligible material strength.

Given its peculiar quasi-satellite dynamical state and the spectroscopic evidence \citep{sharkeyLunarlikeSilicateMaterial2021} pointing to a lunar origin, a plausible scenario is that Kamo'oalewa represents a fragment of lunar crust ejected during a high-energy impact on the Moon. This context could explain its current elongated morphology, which was subsequently preserved during its dynamical evolution within the Earth-Moon system.

\begin{acknowledgements}
Part of this work was supported by \emph{ESO}, project number Ts~17/2--1. JH was supported by the Czech Science Foundation (grant 25-16789S). MD is supported by CNES and is Leverhulme Visiting Professor at the University of Leicester with financial support from the Leverhulme Trust (UK). 
\end{acknowledgements}

%
   \bibliographystyle{aa} 
   \bibliography{zotero.bib,refsMod.bib,mybib}

\appendix

\section{Supplementary Analysis}
\label{sec:suppAnal}
\begin{figure} [htbp]
\centering
\includegraphics[width=\linewidth]{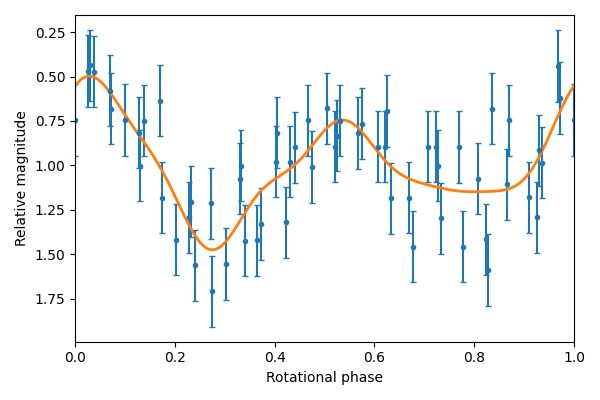}
\caption{Example of a model fit to the observed lightcurve. Blue points with error bars correspond to the photometric measurements, assuming a 0.23 magnitude error, while the orange line shows the synthetic lightcurve derived from the shape model. Data are plotted as a function of rotational phase, and magnitudes are given as relative values.}
\label{fig:fit2lc}
\end{figure}

Information about the photometric uncertainties is not available from the AstDys repository. However, two different approaches can be used to estimate these uncertainties \citep[see, e.g., and references therein]{Hanus2011}. The first is to measure the RMS of the convex inversion fit, which we find to be 0.21 magnitudes, roughly reflecting the noise in the data. The second is to fit a Fourier curve to each lightcurve and compute the RMS. This yields values between 0.10 and 0.32 magnitudes, with an average of 0.23 magnitudes, which is very similar to the value obtained using the first method. Figure~\ref{fig:fit2lc} shows an example of the convex inversion model fit to a lightcurve.

\section{Supplementary figure}

\begin{figure}[h]
   \centering
   \includegraphics[width=0.48\textwidth]{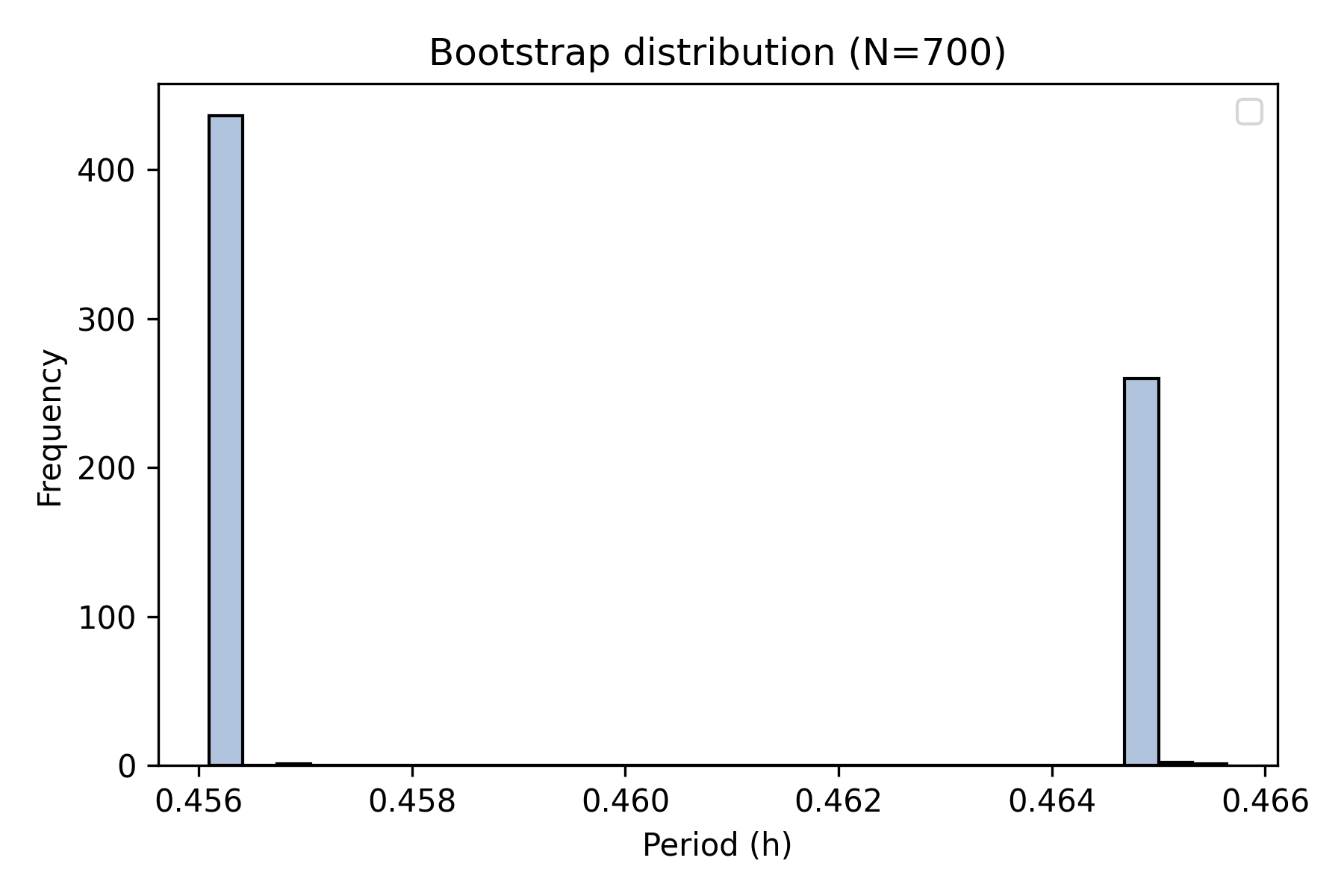}
   \caption{Bootstrap resampling of the photometric dataset. We show a histogram of the best-fitting periods -- the two previously derived sidereal periods clearly dominate. The 0.456~h period is more common -- in $\sim$69\% cases. }
   \label{fig:bootstrap}
\end{figure}

\section{Shape model projections}
\label{sec:shape}

\begin{figure} [htbp]
\centering
\includegraphics[width=1\linewidth, trim=6cm 0cm 6cm 0cm, clip]{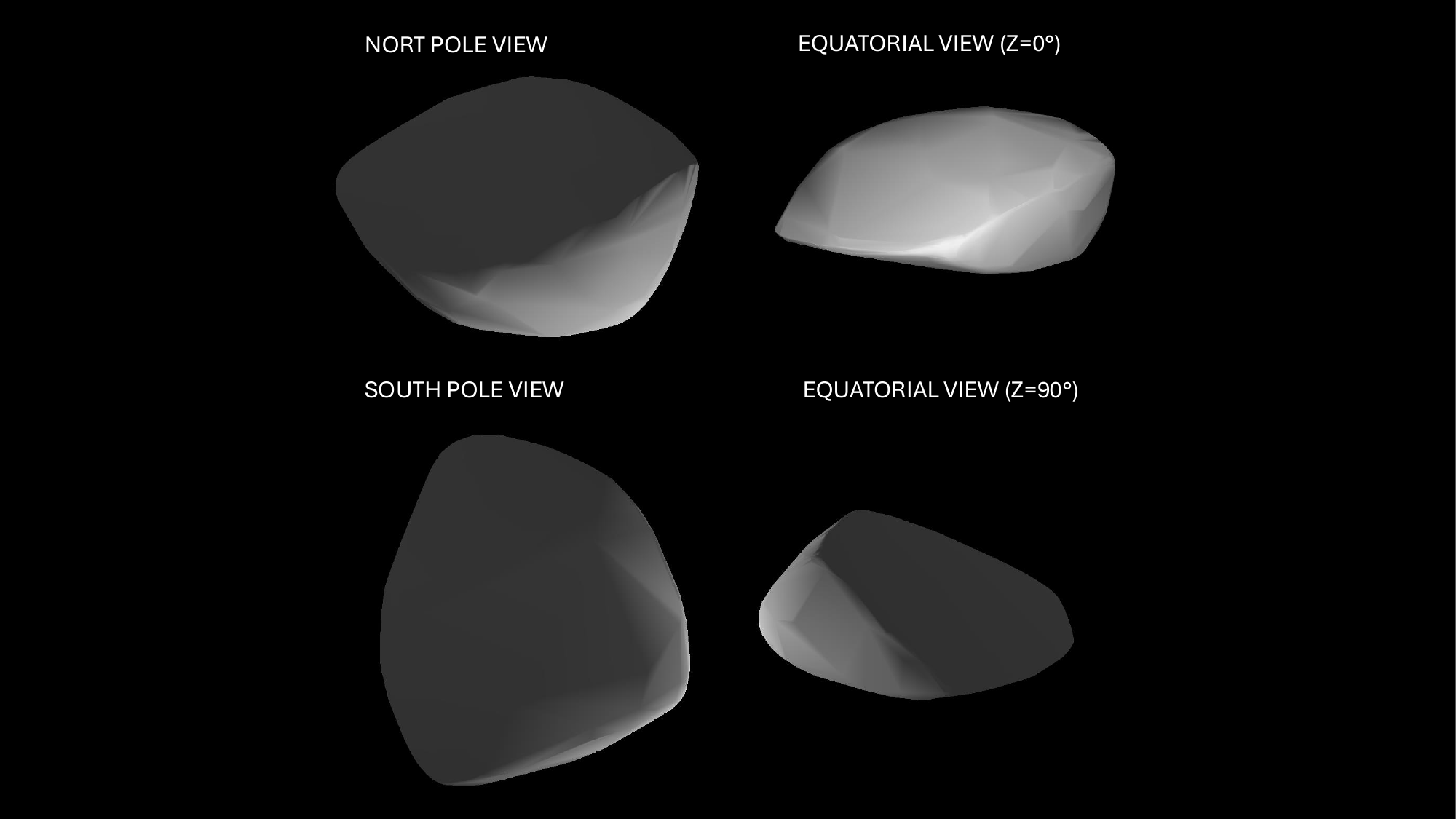}
\caption{Convex shape model corresponding to solution 2 for the 0.465 h period of Kamo'oalewa.}
\label{fig:shape2}
\end{figure}

\begin{figure} [htbp]
\centering
\includegraphics[width=1\linewidth, trim=6cm 0cm 6cm 0cm, clip]{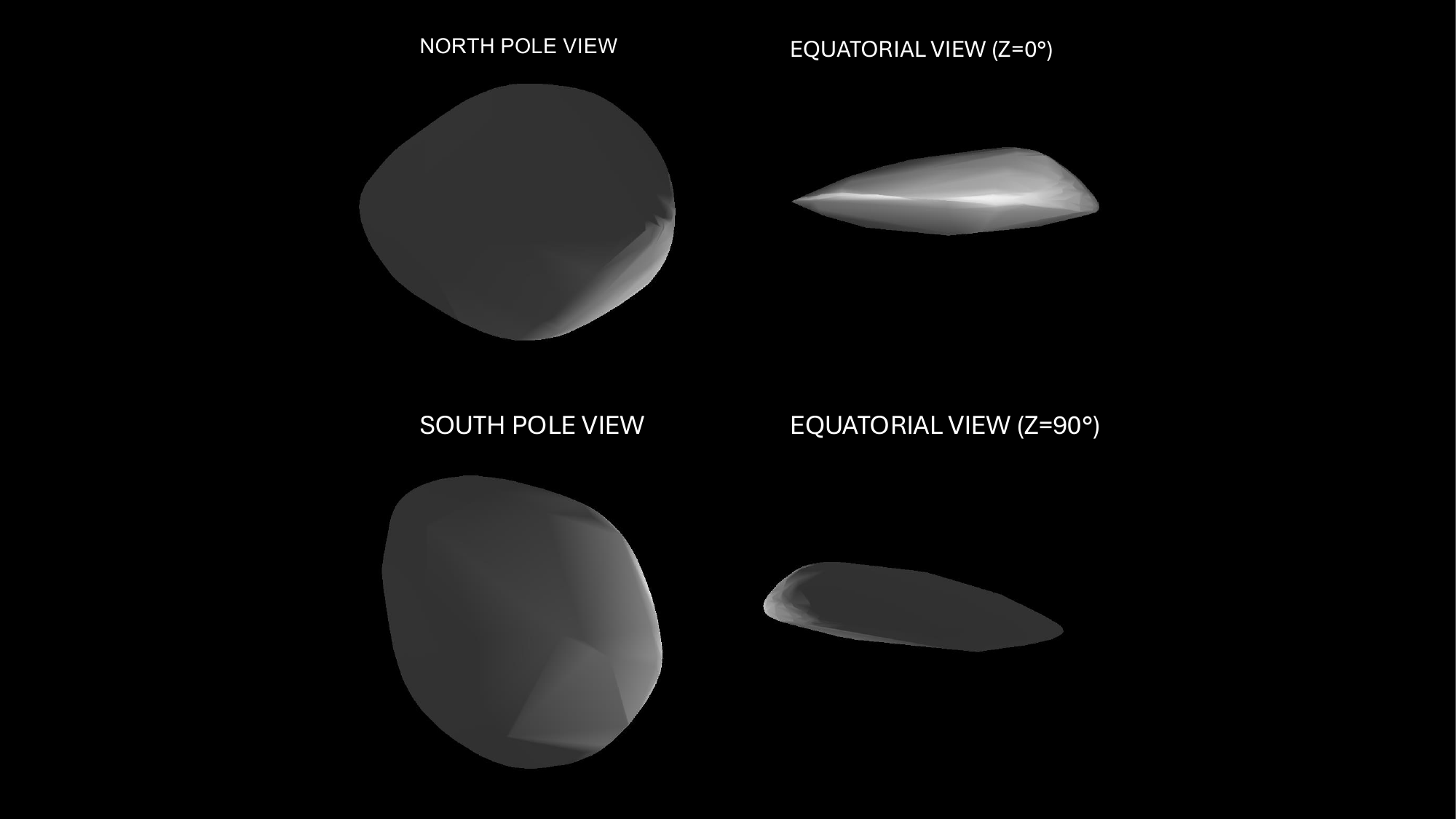}
\caption{Convex shape model corresponding to solution 3 for the 0.465 h period of Kamo'oalewa.}
\label{fig:shape1}
\end{figure}

\begin{figure} [htbp]
\centering
\includegraphics[width=1\linewidth, trim=6cm 0cm 6cm 0cm, clip]{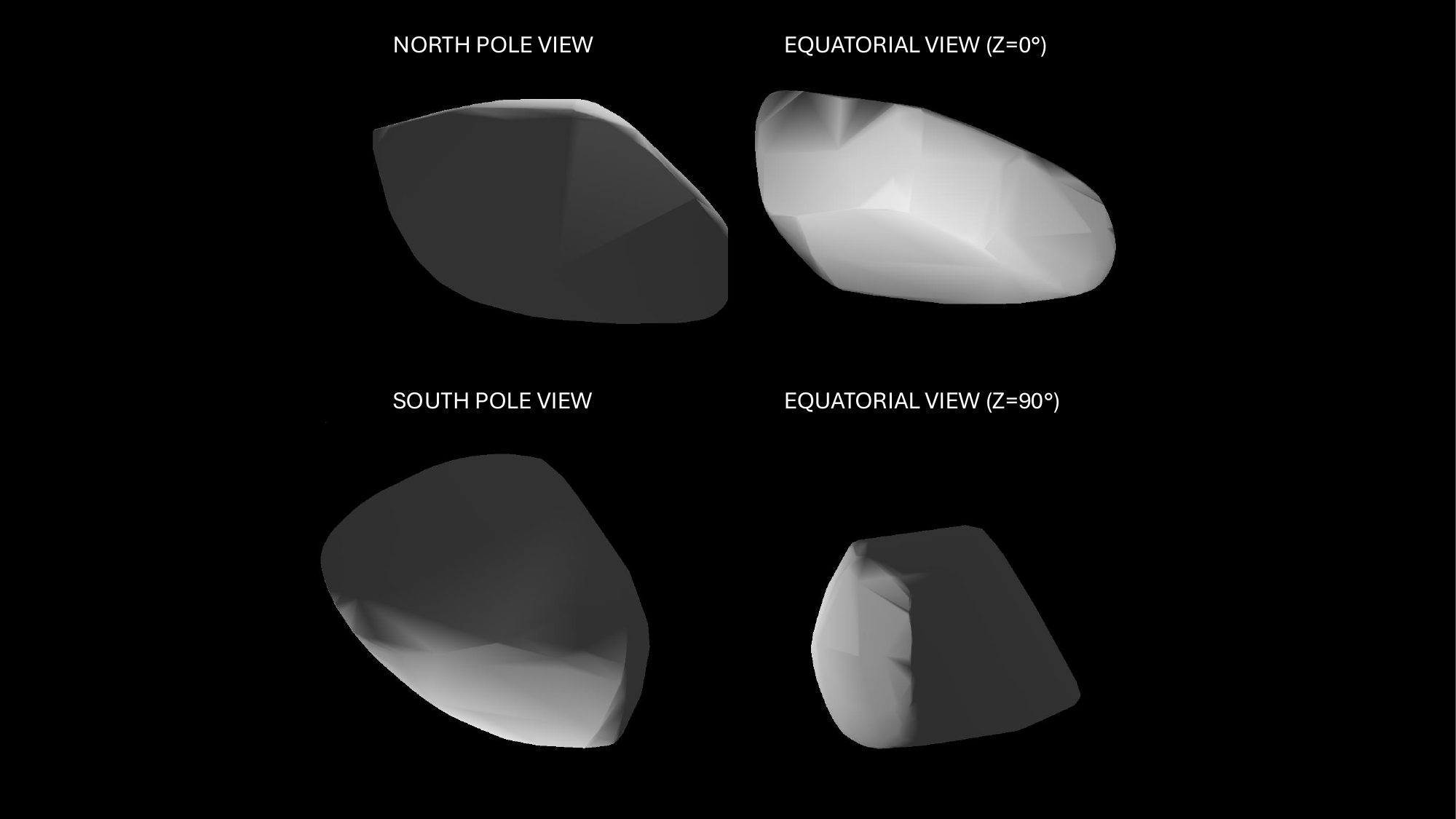}
\caption{Convex shape model corresponding to solution 1 for the 0.456 h period of Kamo'oalewa.}
\label{fig:shape2}
\end{figure}

\begin{figure} [htbp]
\centering
\includegraphics[width=1\linewidth, trim=6cm 0cm 6cm 0cm, clip]{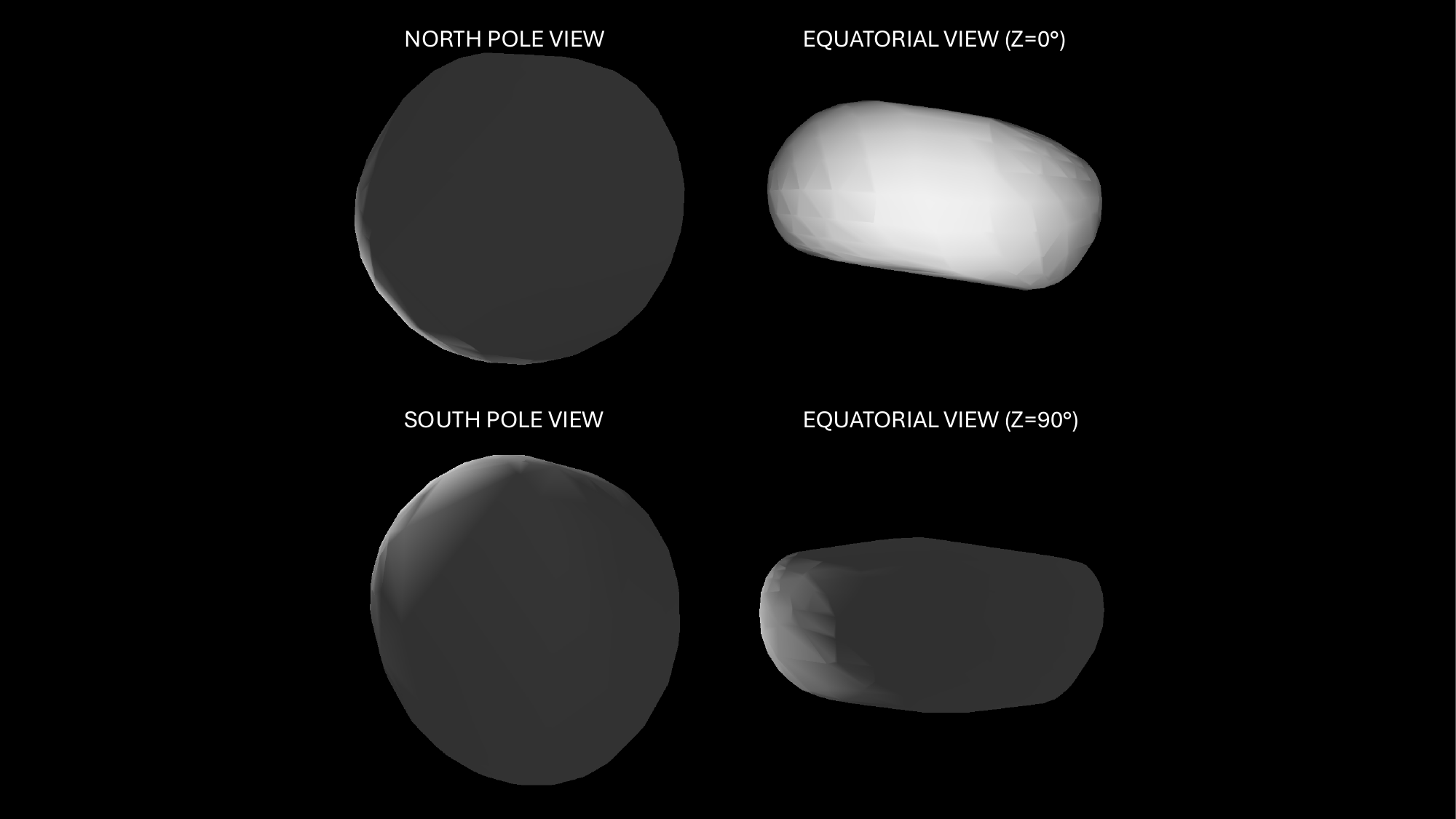}
\caption{Convex shape model corresponding to solution 2 for the 0.456 h period of Kamo'oalewa.}
\label{fig:shape2}
\end{figure}

\begin{figure} [htbp]
\centering
\includegraphics[width=1\linewidth, trim=6cm 0cm 6cm 0cm, clip]{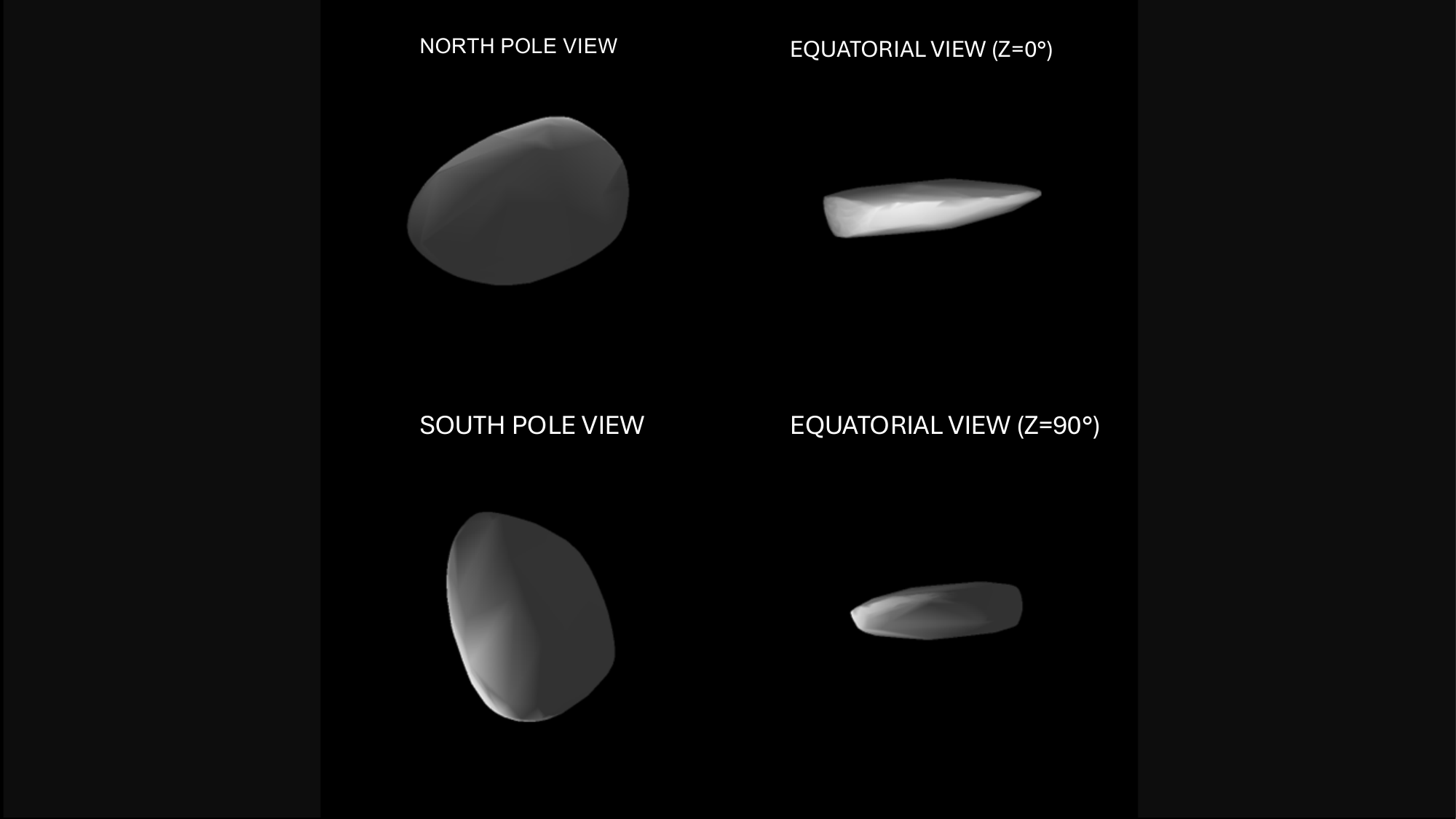}
\caption{Convex shape model corresponding to solution 3 for the 0.456 h period of Kamo'oalewa.}
\label{fig:shape2}
\end{figure}

\end{document}